\documentclass[right,12pt]{amsart}
\usepackage{graphicx}
\usepackage {amssymb}
\usepackage{amsmath}
\begin{document}
\newtheorem{thm}{Theorem}%[section]
 \newtheorem{lem}{Lemma}%[section]
 \newtheorem{prop}{Proposition}[section]
 \newtheorem{defn}{Definition}[section]
  \newtheorem{rem}{Remark}%[section]
\newtheorem{cor}{Corollary}
% \newtheorem{lem}[thm]{Lemma}
% \newtheorem{prop}[thm]{Proposition}

 %\newtheorem{defn}[thm]{Definition}

 %\newtheorem{rem}[thm]{Remark}
% \numberwithin{equation}{section}
%\title{On systems of ODEs associated with the full compressible Euler
%equations in any space dimension}
%\title{A criterium for the gradient
%catastrophe for the non-isentropic gas dynamics  equations%\thanks{Grants or other notes
\title[A criterion of
singularity formation]{A criterion of singularity formation for the
non-isentropic gas dynamics equations}

%for the non-isentropic  gas dynamics

\author{Olga S. Rozanova}

\address{%O.S. Rozanova\at
              Moscow State University, Moscow, 119991, Russia}%\\
\email
{rozanova@mech.math.msu.su}           %  \\


%\maketitle

%\thanks{Supported by grant of DFG  RUS 436
%113/823/0-1}

%\subjclass{35Q35}

%\keywords{gas dynamics equations, the Cauchy problem, gradient
%catastrophe}

%%% ----------------------------------------------------------------------

\begin{abstract}
For the 1D non-isentropic polytropic gas dynamics equations we find
sufficient and necessary conditions for blow up of derivatives in
the terms of  Cauchy data. In particular, the method allows to
determine exact class of initial data corresponding to globally
smooth in time solution.
\end{abstract}

%%% ----------------------------------------------------------------------
\maketitle
%%% -----------------------------

%\vskip2cm

%2010 Mathematical Subject Classification: 35L65, 76N15, 35L67. Key
%Words: Conservation laws, compressible Euler equations, singularity
%formation, large data, exact solutions.

\section{Preliminaries}\label{0}

We consider the system of non-isentropic  gas dynamics equations for
unknown functions $\rho,  v, S, p$ (density,
 velocity, entropy and pressure), namely
\begin{equation}
 \rho (\partial_t v+ v\partial_x v)+\partial_x p= 0,\label{(0.1)}
 \end{equation}
 \begin{equation}
 \partial_t \rho + \partial_x (\rho v)=0,\label{(0.2)}
 \end{equation}
 \begin{equation}
 \partial_t S +v\partial_x S=0.\label{(0.3)}
\end{equation}
   The functions depend on
time $t\ge 0$ and on point $x\in {\mathbb R}.$

The state equation is
\begin{equation}
p=\frac{1}{\gamma}\rho^\gamma e^S,\label{(0.4)}
\end{equation}
 where $\gamma>1$ is the adiabatic
exponent.

%,\cite{Majda}.

%$$(\rho(t,x),\,V(t,x),\,S(t,x))\in
%C^k([0,T);{\mathbb H}^m({\mathbb
%R})),\,m>\frac{n}{2}+k,\,T\le\infty,$$ provided
%$(\rho(0,x),\,V(0,x),\,S(0,x))\in {\mathbb H}^m({\mathbb R})$
%\cite{Kato}.

For classical solutions equations \eqref{(0.2)} -- \eqref{(0.4)}
imply
\begin{equation}
 \partial_t p + v \partial_x p +\gamma \,p\,\partial_x v=0.\label{(0.5)}
 \end{equation}

We consider the Cauchy problem for \eqref{(0.1)}, \eqref{(0.2)},
\eqref{(0.5)} with the data
\begin{equation}
\rho(0,x)=\rho_0(x)>0, \quad {v}(0,x)={ v}_0(x),\quad
p(0,x)=p_0(x)>0.\label{(0.6)}
\end{equation}

%According to \cite{Ovs}, we call the movement of gas {\it smooth}
%inside some volume $\Omega \subset {\mathbb R}$, if the components
%of solution ${V}, \rho, p$ are of the class $C^1(\Omega\times
%{\mathbb R}_+)$ almost everywhere.

System \eqref{(0.1)} -- \eqref{(0.3)} is symmetric hyperbolic,
therefore, it has a local in $t$ solution as smooth as initial data
\cite{Kato}. If we require the initial data \eqref{(0.6)} to be of
class $C^k({\mathbb R}^n),\, k>1,$ then the solution keeps this
smoothness till it is bounded together with its first derivatives
\cite{Majda},\cite{Dafermos}. Further we set $k=1.$

%It is known that $\rho(t,x)$ (and therefore $p(t,x)$) is strictly
%positive if the initial density is separated from zero, moreover,
%the component of solution are bounded until the derivatives are
%bounded \cite{RozhYan}.

%a classical solution of \eqref{(0.1)}, \eqref{(0.2)}, \eqref{(0.5)}
%is separated from

 It is well known that the
solutions of the systems of gas dynamics equations, being arbitrary
smooth initially, can generate  unboundedness of first derivatives
within a finite time. This phenomenon is called the gradient
catastrophe.

However, the sufficient and necessary conditions for the gradient
catastrophe in terms of initial data were known only for the
isentropic one-dimensional flow, where one can write the system of
two quasilinear equations in  Riemannian invariants. For the
non-isentropic case quite numerous but only partial results were
obtained. The nonlinear capacity method gives  sufficient conditions
for the loss of smoothness \cite{Pokhozhaev}. There are attempts to
adapt the characteristics method, but the criterium of the gradient
catastrophe was not attained on the way \cite{LiuFagui},
\cite{Zhao_1989}, \cite{Zhu_1996}.

Recently  some progress in the study the non-isentropic gas dynamics
equation has been made. Namely, based on the Lagrangian form of the
full Euler equations in \cite{Chen_2013} the authors reduced
non-isentropic gas dynamics equations  to a special form that allows
to study the system by analogy with the Riemann invariants under
additional assumptions. The method gives a possibility to find
conditions for the singularity formation and to present several new
examples of shock-free solutions, which demonstrate a large variety
of behaviors \cite{Chen_2011}, \cite{Chen_2014}, \cite{Chen_2015},
\cite{Zheng_2016}. However, the complete picture of the finite time
shock formation from smooth initial data was not achieved.

The result of this paper is obtained by a very classical method. We
consider the augmented system that consists of the equations for the
first derivatives of solution together with the initial system of
gas dynamics. This system was introduced in \cite{Courant_Lax} and
was used there to prove a local smooth solvability of the Cauchy
problem. Further, in \cite{RozhYan} it was noticed that this system
can always be written in  Riemann invariants.  To obtain this system
of  Riemann invariants for the specific case of the gas dynamics
equations one can perform standard but thorough computations. For
the case of gas dynamics this system can be in some sense decoupled.
This is the key point that allows us to reduce the problem to
studying an autonomous system of thee  ordinary differential
equations that can be integrated.

The paper is organized as follows. In Sec.\ref{main_theorem} we
formulate the main results. In Sec.\ref{isentropic} we recall the
criterium of the singularity formation for the isentropic gas
dynamics and show that this criterium is a particular case of our
main theorem. In Sec.\ref{general} we prove the theorem in the
general case.

\section{Main theorem}\label{main_theorem}
%: the criterium of the gradient catastrophe in one-dimensional case

Let us denote $R_1(x)=v_0',$ $R_2(x)=\frac{ p_0'}{\sqrt{\gamma\rho_0
p_0}},$ $b(x)=-\frac{\gamma-1}{2}+ K(x),$ $K(x)=\frac{\gamma p_0
\rho_0'}{2\rho_0 p_0'}-\frac{1}{2}$.

\begin{thm}\label{mt}
%(Criterium of the gradient catastrophe)
 Suppose initial data \eqref{(0.6)} to be
of class $C^1({\mathbb R})$. The solution to \eqref{(0.1)},
\eqref{(0.2)}, \eqref{(0.5)} keeps smoothness for all $t\ge 0$ if
and only if for every point $x\in {\mathbb R}$ one of the following
set of inequalities holds:
\begin{itemize}\item
\begin{equation}\label{cond_smoothb+1}
b\ge 0,
%\min\limits_{x\in{\mathbb R}} b(x)>0
%\quad \mbox{\it and}
\quad  R_1\ge 0,
\end{equation}
\item
\begin{equation}\label{cond_smoothb+2}
b > 0,
%\min\limits_{x\in{\mathbb R}} b(x)>0
\quad R_2\ne 0,
\end{equation}
\item
\begin{equation}\label{cond_smoothb-}
b< 0, \quad R_1\ge 0,\quad  R_1^2+\frac{2 b}{\gamma-1} R_2^2\ge 0,
%\min\limits_{x\in{\mathbb R}} b(x)>0
\end{equation}
\item
\begin{equation}\label{cond_smoothK_inf}
R_2=0, \quad R_1\ge 0.
%\min\limits_{x\in{\mathbb R}} b(x)>0
\end{equation}
\end{itemize}
For all others initial data the derivatives of solution become
unbounded within a finite time
 $T>0$.
\end{thm}

Away from vacuum, the development of singularity in the smooth
solution to the compressible Euler system is due to
unboundedness of derivatives \cite{Lax}. % \cite{Liu}. %see also
%\cite{Smoller}).

Let us remark that the density and pressure taking part of the
smooth solution cannot vanish before the singularity formation.
Indeed, if the derivatives of smooth solution are bounded, then
\eqref{(0.2)}, \eqref{(0.5)}, \eqref{(0.6)} imply
$\rho(t,x)>\phi(t)>0$, $p(t,x)>\psi(t)>0$.

If the derivatives of solutions are bounded for all $t>0$, then the
solution keeps smoothness globally in $t$.

\section{Isentropic case}\label{isentropic}

In the isentropic case $S= const$ the system \eqref{(0.1)},
\eqref{(0.2)}, \eqref{(0.3)} consists of two equations and therefore
it can be written in Riemann invariants.

The criterium of gradient catastrophe for the system of two Riemann
invariants is known since \cite{Borovikov} (see also
\cite{RozhYan}). Let us recall this theorem and the proof.
\begin{thm}{\rm \cite{Borovikov}}\label{rith} Consider the system
\begin{equation}\label{ri}
\partial_t r_k + \xi_k\partial_x r_k=0,\quad k=1,2,
\end{equation}
subject to the Cauchy data
\begin{equation*}
\label{cd} r_k(x,0)=r_k^0(x)\in C^1(\mathbb{R}), \quad k=1,2.
\end{equation*}

Let $r_k=r_k(t,x)$, $\xi_k=\xi_k(r_1,r_2)$ be differentiable
functions and
\begin{equation*}\label{cond_ri}
\frac{\partial \xi_k}{\partial  r_k}>0,\quad k=1,2.
\end{equation*}
If
\begin{equation*}\label{crit_cat}
\min\limits_x \min\limits_k \frac{ d r_k^0(x)}{dx} <0,
\end{equation*}
then the derivatives of the solution become unbounded for some
$t=T>0$. Otherwise, if
\begin{equation*}\label{crit_smooth}
\min\limits_x \min\limits_k \frac{ d r_k^0(x)}{dx} \ge 0,
\end{equation*}
then the solution keeps smoothness for all $t>0.$
\end{thm}

Let us recall the idea of proof \cite{Borovikov} (see also
\cite{RozhYan}, Ch.1, Sec.10.2). Differentiating  \eqref{ri} with
respect to $x$ we get
\begin{equation}\label{ri_c1}
\partial_t p_1 + \xi_1\partial_x p_1= - \frac{\partial \xi_1}{\partial r_1} p_1^2
 -\frac{\partial \xi_1}{\partial r_2} p_1 p_2,
\end{equation}
\begin{equation}\label{ri_c2}
\partial_t p_2 + \xi_2\partial_x p_2=- \frac{\partial \xi_2}{\partial r_2} p_2^2
 -\frac{\partial \xi_2}{\partial r_1} p_1 p_2,
\end{equation}
where $p_k=\partial_x r_k$. Further, introducing positive functions
\begin{equation*}
\phi_1(r_1,r_2)=\exp\left(\int\limits_0^{r_2} \frac{\partial
\xi_1}{\partial \tilde r_2} \frac{1}{\xi_1-\xi_2}\, d \tilde
r_2\right),
\end{equation*}
\begin{equation*}
\phi_2(r_1,r_2)=\exp\left(\int\limits_0^{r_1} \frac{\partial
\xi_2}{\partial  \tilde r_1} \frac{1}{\xi_2-\xi_1}\, d \tilde
r_1\right),
\end{equation*}
we obtain from \eqref{ri_c1}, \eqref{ri_c2}
\begin{equation*}%\label{ri_c1}
\partial_t(\phi_k p_k) + \xi_k\partial_x(\phi_k p_k)= - \chi_k (\phi_k
p_k)^2,\quad \chi_k= \frac{\partial \xi_k}{\partial r_k}\big/
\phi_k, \quad k=1,2.
\end{equation*}
The conclusion of Theorem \ref{rith} follows immediately. $\Box$

The Riemann invariants for the isentropic gas dynamics are
$$r_{1,2}=v\mp \int\limits_0^\rho \frac{c(\tilde \rho)}{\tilde\rho} d
\rho,$$ where $c(\rho)=\sqrt{p_\rho}$ \cite{RozhYan}.  Computations
show that if we set $S=0$, then $$r_{1,2}=v\mp \frac{2}{\gamma-1}
\rho^\frac{\gamma -1}{2},$$ and the system in the variables $r_1,$
$r_2$ has the form
\begin{equation}\label{ri_gd}
\partial_t r_1 + (\alpha r_1 + \beta r_2)\partial_x r_1=0,\quad \partial_t r_2 +
 (\beta r_1 + \alpha r_2)\partial_x r_2=0,
\end{equation}
with $\alpha=\frac12 + \frac{\gamma-1}{4}>0$, $\beta=\frac12 -
\frac{\gamma-1}{4}$. Thus, in the case of \eqref{ri_gd} Theorem
\ref{rith} implies the following corollary.
\begin{cor}\label{cor_ri}
The Cauchy problem \eqref{(0.6)} in the isentropic case $S=const$
has a globally smooth solution if and only if
\begin{equation} \label{smooth_igd}
\min\limits_{x} (v'_0\mp \rho_0^{\frac{\gamma-3}{2}}\rho'_0)\ge 0
\end{equation}
Otherwise, the derivatives of solution go to infinity within a
finite time $T>0$.
\end{cor}

It can be readily shown that this result follows from Theorem
\ref{mt} as well. Indeed, if $p=\frac{1}{\gamma}\rho^\gamma,$ then
$K(x)=0,$ $b(x)<0$. Condition \eqref{cond_smoothb-} implies
\begin{equation*}
\min\limits_{x}((v'_0)^2-\rho^{\gamma-3} (\rho_0')^2) \ge 0
\quad\mbox{\rm and}  \quad \min\limits_{x} v'_0>0,
\end{equation*}
 it can be reduced to
\eqref{smooth_igd}.

\begin{rem} The criterium of the singularity formation for the 1D
isentropic gas dynamics can be obtained by means of different tools,
see e.g. \cite{Chen_2014}.
\end{rem}

\section{General case}\label{general}

In the non-isentropic case the system of gas dynamics cannot be
written in  Riemann invariants. Nevertheless, the augmented system
that includes the components of solutions together with their first
derivatives can always be written in the Riemann invariants
\cite{RozhYan}. This fact was known, but rarely used, since the
resulting system is nonhomogeneous and it seems there is no hope to
analyze it. We are going to show that for the case of the gas
dynamics this system can be written in a reasonable form.

\subsection{Augmented system and its Riemann invariants}
Let us recall the method of obtaining  the system of  Riemann
invariants for any hyperbolic system \cite{RozhYan}, Ch.1, Sec.4.3.
Assume that the strictly hyperbolic system of $n$ equations for the
vector-function $u(t,x)=(u_1,\dots, u_n)$
\begin{equation}\label{main}
\partial u + A(u) \partial u =0,
\end{equation}
where $A(u)$ is a matrix with real different eigenvalues, is written
in the characteristic form
\begin{equation}\label{char_form}
l^k\left[\frac{\partial u}{\partial t}+ \xi_k \frac{\partial
u}{\partial x}\right]=0,
\end{equation}
where $l^k(u)=(l^k_1,\dots, l^k_n) $, $k=1,\dots,n,$ is a left
eigenvector, $\xi_k(u)$ is the respective eigenvalue. If we
introduce the notation $\frac{\partial u}{\partial x}=p$,
$\frac{\partial u}{\partial t}=q$, we can re-write \eqref{char_form}
as
\begin{equation}\label{char_form1}
l^k\left[q+ \xi_k p\right]=0.
\end{equation}
Differentiating \eqref{char_form1} with respect to $t$ and $x$, and
taking into account the condition $\frac{\partial q}{\partial
x}=\frac{\partial p}{\partial t}$ we obtain
\begin{equation}\label{char_form_q}
l^k\left[\frac{\partial q}{\partial t}+ \xi_k \frac{\partial
q}{\partial x}\right]=\mathfrak{G}_k,
\end{equation}
\begin{equation}\label{char_form_p}
l^k\left[\frac{\partial p}{\partial t}+ \xi_k \frac{\partial
p}{\partial x}\right]=\mathfrak{F}_k,
\end{equation}
where
\begin{equation*}\mathfrak{G}_k=-\sum\limits_{i,j=1}^n\left[(q_i+\xi_k
p_i)\frac{\partial l^k_i}{\partial u_j} q_j+l^k_i\frac{\partial
\xi^k}{\partial u_j} q_j p_i\right], \end{equation*}
\begin{equation*}
\mathfrak{F}_k=-\sum\limits_{i,j=1}^n\left[(q_i+\xi_k
p_i)\frac{\partial l^k_i}{\partial u_j} p_j+l^k_i\frac{\partial
\xi^k}{\partial u_j} p_j p_i\right].\end{equation*}

System \eqref{char_form}, \eqref{char_form_q}, \eqref{char_form_p}
is called the augmented system.

Let us denote ${\mathcal P}_k=l^k p=\sum\limits_{i=1}^n l^k_i p_i.$
Since the matrix $\Lambda=l^k_i$ is non-degenerate, we can find
\begin{equation}\label{p}
p_k=\sum\limits_{i=1}^n \lambda^k_i \mathcal{P}_i,
\end{equation}
 where
$\lambda^k_i$ are the components of $\Lambda^{-1}.$ Further, from
\eqref{char_form1} we have
\begin{equation}\label{q}
 q_k=-\sum\limits_{i,j=1}^n \lambda^k_i l^i_j\xi_i p_j
=-\sum\limits_{i=1}^n \lambda^k_i\xi_i{\mathcal P}_i.
\end{equation}
Thus, from \eqref{char_form_p} we obtain
\begin{equation}\label{ri_mathcalp}
\frac{\partial {\mathcal P}_k}{\partial t}+ \xi_k \frac{\partial
{\mathcal P}_k}{\partial x}={\bf F}_k,
\end{equation}
where ${\bf F}_k=\mathfrak{F}_k+\sum\limits_{i,j=1}^n p_i
\frac{\partial \,l^k_i}{\partial\, u_j} (q_j+\xi_k p_j)$. System
\eqref{ri_mathcalp} consists of $n$ equations, it can be considered
together with system of $n$ equations
\begin{equation}\label{ri_u}
\frac{\partial u_k}{\partial t}+ \xi_k \frac{\partial u_k}{\partial
x}=\bf{U}_k,
\end{equation}
where ${\bf U}_k=-\sum\limits_{i=1}^n \lambda^k_i\xi_i{\mathcal P}_i
+\xi_k \sum\limits_{i=1}^n \lambda^k_i \mathcal{P}_i.$ Here $p$ and
$q$ in $\bf{F}_k$ and $\bf{U}_k$ can be expressed through $\mathcal
P$ by formulae \eqref{p} and \eqref{q} such that
$$
{\bf F}_k=F^k(u)+\sum\limits_{i=1}^n  F^k_i(u){\mathcal P}_i
+\sum\limits_{i,j=1}^n F^k_{ij}(u) {\mathcal P}_i{\mathcal P}_j,
$$
$$
{\bf U}_k=U^k(u)+\sum\limits_{i=1}^n  U^k_i(u){\mathcal P}_i,\quad
i=1,\dots,n.
$$
${\bf F}_k$ and ${\bf U}_k$ are quadratic and linear polynomials
with respect to ${\mathcal P}$ with coefficients that depend only on
$u$.

Thus, the quadratically nonlinear system of $2n$ equations
\eqref{ri_mathcalp}, \eqref{ri_u} is written in the Riemann
invariants.

\subsection{Gas dynamics equations}

First of all we re-write system \eqref{(0.1)},\eqref{(0.2)},
\eqref{(0.5)} in terms of velocity $v$, specific volume
$\tau=\frac{1}{\rho}$ and pressure $p$. We enumerate the components
of solution as $u_1=v$, $u_2=\tau$, $u_3=p$. The resulting system is

\begin{equation}
 \partial_t u_1+ u_1\partial_x u_1+ u_2 \partial_x u_3= 0,\label{(1)}
 \end{equation}
 \begin{equation}
 \partial_t u_2 + u_1 \partial_x  u_2 - u_2 \partial_x u_1=0,\label{(2)}
 \end{equation}
\begin{equation}
 \partial_t u_3 + u_1 \partial_x u_3 +\gamma \,u_3\,\partial_x
 u_1=0.\label{(5)}
 \end{equation}
 The initial data $u^0(x)=(u_1^0, u_2^0, u_3^0)$ can be expressed
 through \eqref{(0.6)}.

Thus, system \eqref{(1)}-\eqref{(5)} has the form \eqref{main},
where
$$ A(u)=
\left[ \begin {array}{ccc} {u_1}&0&{u_2}\\
\noalign{\medskip}-{ u_2}&{u_1}&0\\ \noalign{\medskip}{
\gamma}\,{u_3}&0&{u_1}
\end {array} \right],
$$
with eigenvalues $$ \xi_1=u_1-\sqrt {\gamma u_3 u_2}, \quad
\xi_2=u_1, \quad \xi_3=u_1+\sqrt {\gamma u_3 u_2},$$ and respective
left eigenvectors
$$l^1=\left[ \begin {array}{ccc} 1&0&-s^{-1}\end {array} \right],$$$$ l^2=\left[ \begin {array}{ccc} 0&s&s^{-1}\end {array}
 \right],$$$$ l^3= \left[ \begin {array}{ccc} 1&0& s^{-1}\end {array}
 \right],
$$
where $s=\sqrt{\gamma u_3/u_2}$. Thus, ${\mathcal P}_1= p_1- p_3/s$,
 ${\mathcal P}_2= s p_2+ p_3/s$,  ${\mathcal P}_3=p_1+ p_3/s$,
 and after computation we obtain \eqref{ri_mathcalp} and \eqref{ri_u} with
 $${\bf F}_1=-\frac{\gamma+1}{4} {\mathcal P}_1^2 +
 \frac{1}{4} {\mathcal P}_1 {\mathcal P}_2-
 \frac{3-\gamma}{4} {\mathcal P}_1 {\mathcal P}_3
- \frac{1}{4} {\mathcal P}_2 {\mathcal P}_3,
  $$
$${\bf F}_2=-\frac{\gamma+1}{4}  {\mathcal P}_2 ( {\mathcal P}_1+{\mathcal P}_3),
  $$
$${\bf F}_3=-\frac{\gamma+1}{4} {\mathcal P}_3^2 +
 \frac{1}{4} {\mathcal P}_1 {\mathcal P}_2-
 \frac{3-\gamma}{4} {\mathcal P}_1 {\mathcal P}_3
- \frac{1}{4} {\mathcal P}_2 {\mathcal P}_3,
  $$

$${\bf U}_1=-{\sqrt{\gamma u_2 u_3}}{\mathcal P}_3, \quad {\bf U}_2=\frac{{u_2}}{{2}}{\mathcal
P}_1+\frac{{u_2}}{{2\gamma  u_3}}{\mathcal P}_3,\quad {\bf U}_3=
-\gamma u_3 {\mathcal P}_1,
  $$
  subject to initial data
 \begin{equation}\label{ri_1d_cd}
{ {\mathcal P}}_k\Big|_{t=0}={\mathcal P}_k(u^0(x)),\quad { {u
}}_k\Big|_{t=0}={ u}_k^0(x),\quad k=1,2,3,
\end{equation}
We can see that the left hand sides ${\bf F}_k={\bf F}_k({\mathcal
P})$  do not depend of $u$. This is the key point that allows us to
prove the theorem.

Now let us consider all component of solution ${\mathcal P}, u$ as
functions of $t$ and $\bar x=(x_1, x_2, x_3)$. We denote these new
functions as $\bar {\mathcal P}, \bar u$  and change the system
\eqref{ri_mathcalp}, \eqref{ri_u} to
\begin{equation}\label{ri_Q3d}
\frac{\partial {\bar {\mathcal P}}_k}{\partial t}+
\sum\limits_{i=1}^3\xi_i(\bar u) \frac{\partial {\bar {\mathcal
P}}_k}{\partial x_i}={\bf F}_k(\bar {\mathcal P}),
\end{equation}

\begin{equation}\label{ri_u3d}
\frac{\partial \bar u_k}{\partial t}+  \sum\limits_{i=1}^3\xi_i
(\bar u) \frac{\partial \bar u_k}{\partial x_i}={\bf U}_k (\bar
{\mathcal P},\bar u).
\end{equation}

Let us set the following Cauchy problem for \eqref{ri_Q3d},
\eqref{ri_u3d}:
\begin{equation}\label{ri_3d_cd}
{\bar {\mathcal P}}_i(0,x_1, x_2, x_3)={\mathcal P}^0_i(x_i),\quad
{\bar { u}}_i(0,x_1, x_2, x_3)={ u}^0_i(x_i),\quad i=1,2,3.
\end{equation}

This allows us to study the solution to the Cauchy
problem\eqref{ri_Q3d}, \eqref{ri_u3d}, \eqref{ri_3d_cd} along the
rays
\begin{equation}\label{bichar3d}
\frac{d x_k}{dt}=\xi_k(\bar u),\quad x_k(0)=x_k^0, \quad i=1,2,3,
\end{equation}
 which are lines with coordinates $(x_1(t),
x_2(t), x_3(t))$ in the 3d space. Along the rays \eqref{bichar3d}
the system \eqref{ri_Q3d}, \eqref{ri_u3d} takes the form
\begin{equation*}\label{char3d_xi_u}
\frac{d \bar u_k}{dt}= {\bf U}_k(\bar{\mathcal P},\bar u),
\end{equation*}
\begin{equation}\label{char3d_Q}
\frac{d \bar{ \mathcal P}_k}{dt}={\bf F}_k (\bar{\mathcal P}),\quad
k=1,2,3,
\end{equation}
moreover, the ODE system \eqref{char3d_Q} can be solved separately
subject to initial data $\bar {\mathcal P}_k(0,x_1^0, x_2^0,
x_3^0).$

 Thus, we can find the criterion for a
finite time blowup for the functions $\bar {\mathcal
P}_k(t,x_1^0,x_2^0,x_3^0)$ in dependence on initial data.

 If
we succeed in finding of solution of \eqref{ri_Q3d}, \eqref{ri_u3d},
\eqref{ri_3d_cd}, then we can get the solution to the Cauchy problem
\eqref{ri_mathcalp}, \eqref{ri_u}, \eqref{ri_1d_cd}

as
$${\mathcal P}_k(t,x)=\bar {\mathcal
P}_k(t,x_1,x_2,x_3)\Big|_{x_k=x,\, x_j=x_0,\, j\ne k},$$
$$u_k(t,x)=\bar u_k(t,x_1,x_2,x_3)\Big|_{x_k=x,\, x_j=x_0,\, j\ne k}.$$

%$${\mathcal P}_k(t,x)=\bar {\mathcal
%P}_k(t,x_1,x_2,x_3)\Big|_{x_k=x,\, x_j=0,\, j\ne k},$$
%$$u_k(t,x)=\bar u_k(t,x_1,x_2,x_3)\Big|_{x_k=x,\, x_j=0,\, j\ne k}.$$

Thus, if $\bar {\mathcal P}_k(t,x_1,x_2,x_3)$ blows up (does not
blow up) along the ray  starting from the point $(x_0,x_0,x_0)$, so
does ${\mathcal P}_k(t,x)$ along every characteristic curves $
\frac{d x_k}{dt}=\xi_k(u),\quad i=1,2,3$, starting from the point
$x_0$, its projections to the planes $(x_2=x_0, x_3=x_0)$,
$(x_1=x_0, x_3=x_0)$ and $(x_1=x_0, x_2=x_0)$, respectively.

%Thus, the system of  ODE equations for characteristic curves for the
%system \eqref{ri_Q3d}, \eqref{ri_u3d} in the space of variables
%$x_k(t)$ %$\eta \in \mathbb R$ % parameterized by a
% is
%\begin{equation*}\label{char3d_xi_u}
%\frac{d x_k}{dt}=\xi_k(\bar u),\quad \frac{d \bar u_k}{dt}= {\bf
%U}_k({\mathcal P},\bar u),
%\end{equation*}
%\begin{equation}\label{char3d_Q}
%\frac{d {\mathcal P}_k}{dt}={\bf F}_k ({\mathcal P}),\quad k=1,2,3.
%\end{equation}
%The system \eqref{char3d_Q} can be solved separately and

The finite time blow up of ${\mathcal P}_k(t)$ implies the finite
time gradient catastrophe of the first derivatives of solution $u$.

Thus, we will concentrate on the behavior of the function $\bar
{\mathcal P}_k(t,x_1,x_2,x_3)$, which is governing by system
\eqref{char3d_Q}. For the sake of simplicity we denote $\bar
{\mathcal P}$ as ${\mathcal P}$.

First of all \eqref{char3d_Q} implies
$$
\frac{d({\mathcal P}_1-{\mathcal P}_3)}{d {\mathcal
P}_2}=\frac{{\mathcal P}_1-{\mathcal P}_3}{{\mathcal P}_2},
$$
therefore
\begin{equation}\label{K}{\mathcal
P}_2=K ({\mathcal P}_1-{\mathcal P}_3), \quad K=\rm const,
\end{equation}
and \eqref{char3d_Q} can be reduced to two equations:
\begin{equation}\label{char3d_Q1}
\frac{d {\mathcal P}_1}{dt}=-\frac{(\gamma+1)- K}{4} {\mathcal
P}_1^2+\frac{(\gamma-3)-2 K}{4} {\mathcal P}_1 {\mathcal P}_3 +
\frac{K}{4} {\mathcal P}_3^2,
\end{equation}
\begin{equation}\label{char3d_Q3}
\frac{d {\mathcal P}_3}{dt}=-\frac{(\gamma+1)- K}{4} {\mathcal
P}_3^2+\frac{(\gamma-3)-2 K}{4} {\mathcal P}_1 {\mathcal P}_3 +
\frac{K}{4} {\mathcal P}_1^2.
\end{equation}
Let us introduce new variables $R_1=\frac{{\mathcal P}_1+{\mathcal
P}_3}{2}$, $R_2=\frac{{\mathcal P}_3-{\mathcal P}_1}{2}$. Then
\eqref{char3d_Q1} and \eqref{char3d_Q3} result in
\begin{equation}\label{char3d_R1} \frac{d R_1}{dt}=-
R_1^2 + b R_2^2,
\end{equation}
\begin{equation}\label{char3d_R2} \frac{d R_2}{dt}=-\frac{\gamma+1}{2}
R_1 R_2,
\end{equation}
where $b=-\frac{\gamma-1}{2} +K$. System \eqref{char3d_R1},
\eqref{char3d_R2} has the first integral
\begin{equation}\label{char3d_first_int}
R_1^2= C R_2^{\frac{4}{\gamma+1}}-\frac{2 b}{\gamma-1}R_2^2,
\end{equation}
with $$C=\left(R^2_1(x_0)+\frac{2\gamma b}{\gamma-1}
R_2^2(x_0)\right) R_2^{-\frac{4}{\gamma+1}}(x_0).
$$
Thus, if $b> 0$, then the trajectories on the phase plane are closed
and tend to the origin always except for the case $R_1(0)<0,$
$R_2(x_0)=0$. If $b<0$, then the trajectories tend to the origin
only for $C\ge 0$ and $R_1(x_0)\ge 0.$ If $b=0$,  system
\eqref{char3d_R1}, \eqref{char3d_R2} decompose and the solution is
bounded if $R_1(x_0)\ge 0$. As follows from \eqref{K}, the infinite
value of the constant $K$ corresponds to the case
$Q_1(x_0)-Q_3(x_0)=0$, $R_2(x_0)=0$. Here  system \eqref{char3d_R1},
\eqref{char3d_R2} also can be reduced to one equation
\eqref{char3d_R1} and the solution is bounded if and only if
$R_1(x_0)\ge 0$.

 Thus, above conditions  imply boundedness of derivatives of solution
for all $t>0$ and therefore the solution keep smoothness for all
$t>0$. They correspond to the cases \eqref{cond_smoothb-},
\eqref{cond_smoothb+1}, \eqref{cond_smoothb+2},
\eqref{cond_smoothK_inf}.

Otherwise, the derivatives go to infinity. Moreover, in all that
cases \eqref{char3d_R2} and \eqref{char3d_first_int} imply that   $
\frac{d R_2}{dt}\sim c_0 |R_2|^\sigma R_2$ for large  $R_2$ with a
positive constant $c_0$ and  $\sigma>0$. This means that $R_2$ (and
$R_1$ as well) goes to infinity within a finite time and the
solution has a gradient catastrophe.

\begin{figure}[h]
\begin{minipage}{0.495\columnwidth}
\centerline{\includegraphics[width=0.7\columnwidth]{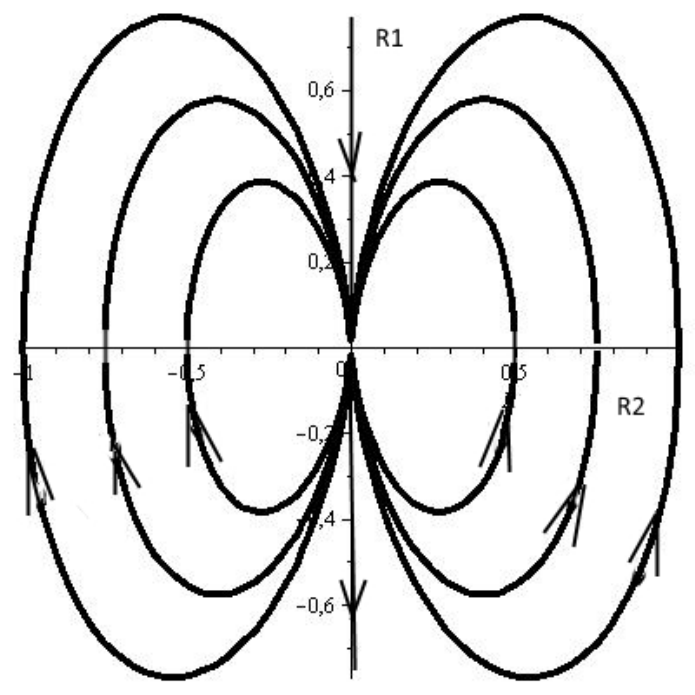}}
\end{minipage}
\begin{minipage}{0.495\columnwidth}
\centerline{\includegraphics[width=0.7\columnwidth]{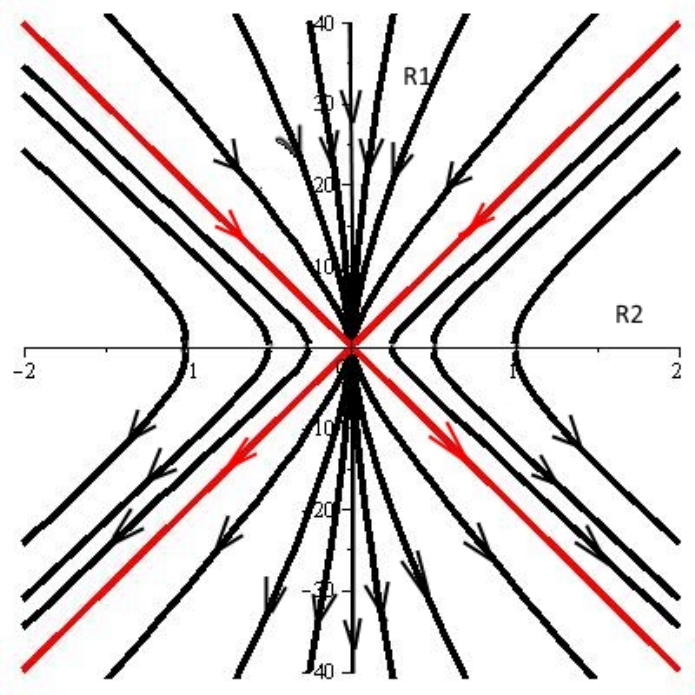}}
\end{minipage}
\caption{The behavior of the phase curves of system
\eqref{char3d_R1}, \eqref{char3d_R2} for $b>0$ (left) and $b<0$
(right).} \label{Pic1}
\end{figure}
%Fig.\ref{Pic1} illustrates the behavior of the phase curves for
%$b>0$ (left) and $b<0$ (right).

Thus, Theorem \ref{mt} is proved.

\begin{rem} Analysis of conditions
\eqref{cond_smoothb+1}, \eqref{cond_smoothb+2},
\eqref{cond_smoothb-}, \eqref{cond_smoothK_inf} shows that one can
find initial density and pressure such that the solution keeps
global smoothness for any initial velocity. For example,  we can
take $p_0=\exp x$, $\rho_0=\exp k x$, $k=1+\frac{2}{\gamma}$. The
data corresponds to the case \eqref{cond_smoothb+2}, with
$b(x)\equiv 1$, $R_2(x)\ne 0$. Such situation is not possible in the
isentropic case.

Further we note  that the extrema of  pressure plays very important
role. Indeed, in these points $R_2=0$ and if the derivative of
velocity is negative, the points generate singularities of solution.
Comparison with conditions \eqref{smooth_igd} shows that for
isentropic case this property also takes place.
\end{rem}
\begin{rem} Since $S=\ln \frac{p}{\rho^\gamma}$, then
$K(x)=-\frac{1}{2}\frac{S_0'(x)}{\rho_0^{\gamma+1}(\ln p_0(x))'}.$
Thus, \eqref{cond_smoothb+2} implies that if the entropy
monotonically decrease
 and pressure monotonically increase (or vice versa),
 and $\frac{S_0'(x)}{(\ln p_0(x))'}< -
(\gamma-1)\,\rho_0^{\gamma+1}$, then the respective solution is
globally smooth. Let us notice that the monotonic profile of entropy
was used in \cite{Chen_2015} to construct smooth solutions to the
compressible Euler equations.
\end{rem}

\begin{rem} The proof does not require the restrictement
$\gamma>1$. In the isothermal case $\gamma=1$ the only difference
will be in the first integral \eqref{char3d_first_int}, it should be
replaced by
\begin{equation*}
R_1^2= C R_2^2-{2 b}R_2^2 \ln |R_2|,
\end{equation*}
with $$C=\left(R^2_1(x_0)+{2b} \ln | R_2(x_0)|\right) R_2^{-2}(x_0).
$$
The conclusion on Theorem \ref{mt} holds as well.

 The result  can be easily modified for any $\gamma\ne 0$. In
 particular, for $\gamma=-1$ (the Chaplygin gas), where the initial system
 is weakly nonlinear, \eqref{char3d_R1}, \eqref{char3d_R2} imply
 $R_2=R_2(x_0)$, $\frac{d R_1}{dt}=-R_1^2+b R_2^2(x_0)$, $b=1+K(x_0)$.
 For the isentropic case ($K(x)\equiv 0$) the function $R_1$ goes to
 $-\infty$  within a finite time if
 \begin{equation}\label{Chap}
 R_1(x_0)<-|R_2(x_0)|
\end{equation}
for a certain $x_0\in \mathbb R$.
 This means that the strict hyperbolicity fails on the respective
 solution. The singularity that arises in this case comprises a delta-like singularity in the component of density
\cite{Kong}. In particular, our results imply that the sufficient
conditions of singularity formation from smooth initial data found
in \cite{Kong} are far to be exact. Condition \eqref{Chap}
corresponds to hypothesis II from \cite{Kong}.
\end{rem}

\begin{rem} A natural question is whether it is possible to
extend the method to the case of several space variables. The answer
is positive for the hyperbolic system of the form
$$\partial_t u+\sum\limits_{k=1}^n A_k \partial_{x_k}u=0,\quad u= (u_1,\dots, u_n),\quad x\in {\mathbb R}^n,$$
where the matrices $A_k$ have a joint set of left eigenvectors. In
the framework of gas dynamics this will be only in the case $p=0$.
The resulting system contains the multidimensional non-viscous
Burgers equation. The criterium for the singularity formation in the
smooth solution to the Cauchy problem is known for this vectorial
equation \cite{protter}, \cite{AKR}. Therefore here the method does
not give us anything new.
\end{rem}


\begin{thebibliography}{15}


%\bibitem{Nedeljkov} M. Nedeljkov , {\em Higher order shadow waves and delta shock blow up in the Chaplygin
%gas}, J. Diff, Equat. {\bf 256} (2014), 3859-3887.







%\bibitem{Ovs} Ovsyannikov, L.V., \emph{Lectures on foundations of gas dynamics},
%2003.
\bibitem{AKR} { S.Albeverio, A.Korshunova, O.Rozanova},
{ A probabilistic model associated with the pressureless gas
dynamics},\emph{ Bull.Sci.Math.}, {\bf 137}(2013), 902--922.


\bibitem{Borovikov} V.A.Borovikov, An upper bound for the existence time of the smooth solution of a quasi- linear
hyperbolic system, \emph{ Sov. Math., Dokl.} \textbf{ 12}, 1586-1590
(1971).

\bibitem{Chen_2011} G. Chen, Formation of singularity and smooth wave propagation for the non-isentropic compressible Euler equations,
\emph{J. Hyper. Differential Equations,} \textbf{8}(2011), 671--290.




\bibitem{Chen_2013}G. Chen, R. Young, Q. Zhang, Shock formation in the compressible Euler equations and related
systems, \emph{ Journal of Hyperbolic Differential Equations} 10
(2013), 149--172.


\bibitem{Chen_2014} G.Chen, R.Pan, Sh.Zhu, Singularity formation for compressible Euler
equations,  E-print:  arXiv:1408.6775 [math.AP].


\bibitem{Chen_2015} G. Chen, R. Young, Shock-free solutions of the compressible Euler
equations, \emph{ Archive for Rational Mechanics and Analysis}
\textbf{217} (2015), 1265-1293.


\bibitem{Courant_Lax}
Courant, R., Lax, P. On nonlinear partial differential equations
with two independent variables. \emph{Comm. Pure Appl. Math.}
\textbf{2} (1949)Vol. 2,  255--273.


\bibitem{Dafermos}C.M.Dafermos, Hyperbolic Conservation Laws in Continuum Physics
(Grundlehren Der Mathematischen Wissenschaften), 3rd Edition,
Springer, 2010.

\bibitem{Kato} Kato, T., The Cauchy problem for quasilinear
symmetric hyperbolic systems,
\emph{Arch.Ration.Math.Anal.}\,\textbf{58}(1975), 181--205.

\bibitem{Kong}D.-X. Kong, Ch.Wei, { Formation and propagation of singularities in one-dimensional
Chaplygin gas}\emph{ J.Geom.Phys.}
 {\bf 80} (2014) 58-70.


\bibitem{Lax} Lax, P.D., The formation and decay of shock waves,
\emph{Amer.Math.Monthly,} \,\textbf{79} (1972), 227--241.

\bibitem{protter}{ H.A.Levine, M.H.Protter},  {  The breakdown of
solutions of quasilinear first order systems of partial differential
equations}, \emph{ Arch.Rat.Mech.Anal.} {\textbf 95}(1986),
253--267.

%\bibitem{Liu} Liu, T.-P., Development of singularities in the
%nonlinear waves for quasilinear hyperbolic partial differential
%equations, \emph{J.Diff.Equat.}\, \textbf{33} (1979), 92--111.


\bibitem{LiuFagui}
Liu, Fagui, Global smooth resolvability for one-dimensional gas
dynamics systems. \emph{Nonlinear Anal., Theory Methods Appl.}
\,\textbf{36}, No.1(A) (1999), 25-34.


\bibitem{Majda} Majda,A., Compressible fluid flow and systems
of conservation laws in several space variables,
\emph{Appl.Math.Sci.}\, \textbf{53}(1984), 1--159.

\bibitem{RozhYan}  Rozhdestvenskij, B.L.; Yanenko, N.N.
\emph{Systems of quasilinear equations and their applications to gas
dynamics}\, Providence, R.I.:AMS, 1983.



\bibitem{Pokhozhaev} Pokhozhaev, S.I., On hyperbolic system of conservation laws, \emph{Differ.
Equat.,} \,\textbf{39}(8)(2003), 663-673.



\bibitem{Zhao_1989} Y.C. Zhao, A class of global smooth solutions of the one dimensional
gas dynamics system, IMA Series (1989).

\bibitem{Zheng_2016} H. Zheng,
Singularity formation for the compressible Euler equations with
general pressure law,\emph{ Journal of Mathematical Analysis and
Applications},  \textbf{438} (2016),  59--72.

\bibitem{Zhu_1996} Ch. Zhu, Global smooth solution of the nonisentropic gas dynamics
system, \emph{ Proceedings of the Royal Society of Edinburgh,}
\textbf{126A} (1996) 769-775.













\end{thebibliography}
\end{document}